\documentclass[
  ,draft            
  ]
  {aipproc}

\layoutstyle{6x9}

\begin{document}
\newcommand{\etal}{{\rm et~al.}}
\newcommand{\kms}{{km~s$^{-1}$}}

\title{Star Formation At High Redshift}

\author{Chip Kobulnicky}{
  address={Department of Physics \& Astronomy, University of Wyoming,
Laramie, WY 82070, USA}
}



\begin{abstract}

I review the observational characteristics of intermediate-to-high
redshift star forming galaxies, including their star formation rates,
dust extinctions, ISM kinematics, and chemical compositions.  I
present evidence that the mean rate of metal enrichment,
$\Delta{Z}/\Delta{z}$, from $z=$0---3, as determined from nebular
oxygen abundance measurements in star forming galaxies, is 0.15 dex
per redshift unit for galaxies more luminous than $M_B=-20.5$.  This
rate of chemical enrichment is consistent with the chemical rise in
Galactic disk stars.  It is less dramatic than, but perhaps consistent
with, the enrichment rate of 0.18--0.26$\pm$0.07 dex per redshift unit seen in
Damped Ly~$\alpha$ systems, and it is much less than predicted by many
cosmological evolution models.  The high-redshift galaxies observed to
date are the most luminous examples from those epochs, and thus, trace
only the greatest cosmological overdensities. Star formation in the
first 1-2 Gyr appears sufficient to elevate ambient metallicities to
near or above the solar value, implying efficient production and
retention of metals in these densest environments.
 
\end{abstract}

\maketitle


\section{Understanding Star Formation at High Redshift  }

Even those of us in the relatively junior stages of our careers
remember the days when ``high-redshift'' meant $z\sim1$ for galaxies
and $z\sim3$ for a handful of the most distant known quasars.  Less
than 15 years ago, the discovery of a ``normal'' galaxy at a redshift
greater than one was breaking news (e.g, \cite{TG},\cite{Te}), and
even the quasars at $z>4$ numbered only 11 \cite{Turner}. Today, by
contrast, catalogs of \  ``normal'' star forming galaxies at redshifts of
$z\geq3$ contain several thousand entries \cite{Se03}, a testament to
the rapid growth in the fields of star formation and galaxy evolution
at cosmological distances, enabled by advances in astronomical
instrumentation.  The limits of discovery continue to be pushed aside
with reports of star forming galaxies at redshifts exceeding $z=6$
\cite{Tan}\cite{Rhoads2} when the universe was less than 800 Myr old!

In this review I will focus on observational measurements of four
basic parameters of star forming galaxies at redshifts $0.3<z<4$: star
formation rates, extinctions, kinematics of the ISM, and chemical
composition.  Specifying even the average properties for these distant
galaxies is complicated by the use of multiple observational detection
and selection techniques, each with its own inherent biases toward
certain classes of galaxies.  Figure~1 uses a Venn-style schematic to
illustrate the variety and degree of overlap between techniques.  By
far the most prolific approach has been the traditional optical
spectroscopic survey to identify restframe optical and ultraviolet
stellar and interstellar features (e.g., \cite{CFA}\cite{CFRS}
\cite{Col}\cite{Strauss}\cite{Coil},).  More specialized methods
for finding high redshift objects include the continuum imaging using
Lyman dropout approach \cite{Se03} and spectroscopic surveys for Lyman
$\alpha$ emission \cite{Rhoads1}.  These preferentially select, respectively, 
objects
with strong measurable continuum and objects with strong emission
lines having particularly favorable ISM geometries to allow the escape
of Lyman $\alpha$ photons, respectively.  Other approaches employ
multi-frequency radio continuum imaging
\cite{Carilli}\cite{Ivison} or use millimeter-wave measurements to detect dust
emission from the host galaxy \cite{Hughes}\cite{Eales}\cite{Webb}.  These
varied detection techniques complement one another by selecting
different (but perhaps related, and sometimes overlapping) high
redshift galaxy populations.  However, they complicate efforts to
ascribe \ ``average'' properties to these distant objects.  For the
purposes of this review, I will focus on the optically selected
objects and simply caution that these are unlikely to be
representative of the mean extinctions, star formation rates, or
chemical composition of radio-selected or submillimeter-selected samples.
Furthermore, I address only galaxies powered by star formation
and do not consider objects with active nuclei. 

\begin{figure}
  \includegraphics[height=4.5in,angle=0]{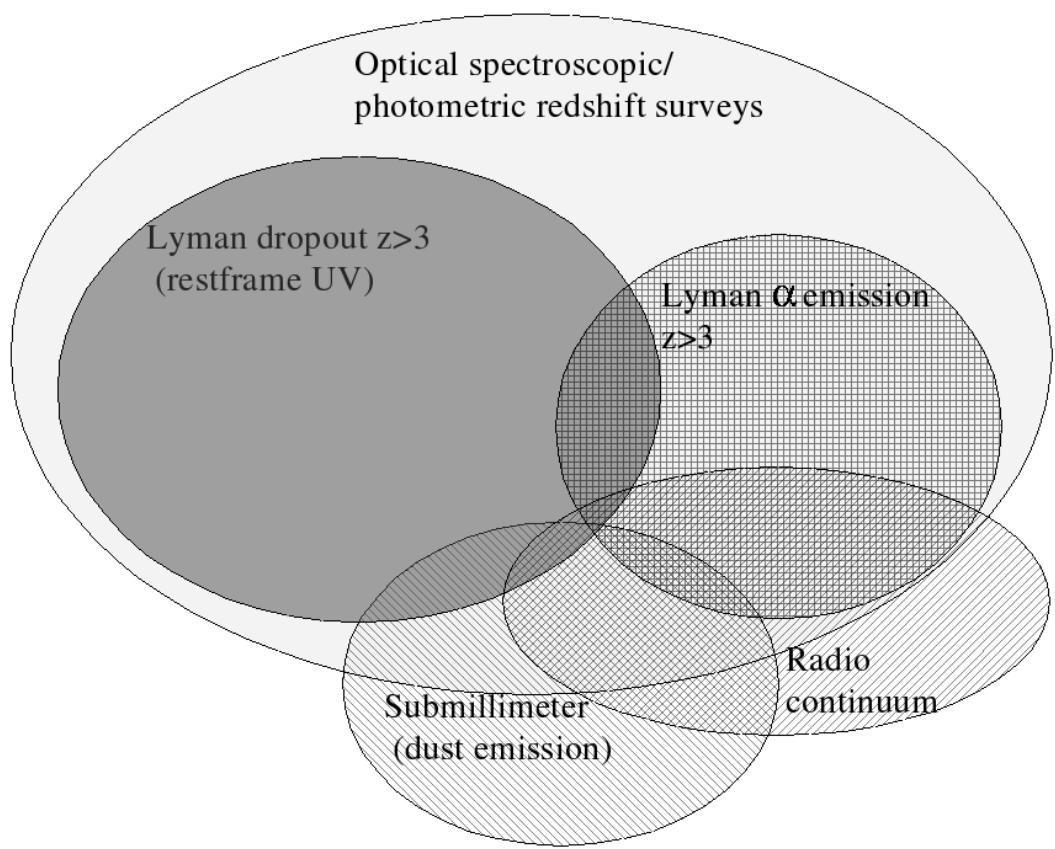}
 \caption{Venn-style schematic showing the efficacy of widely used 
techniques in the detection of high redshift galaxies.  }
\end{figure}

The environment governing star formation in galaxies at high redshift
is shaped by the prevailing universal conditions imposed by cosmology.
Distant galaxies reside in a younger, smaller, more gas-rich,
more metal-poor universe where the supply of the raw material for star
formation is abundant.  Figure~2 shows a plot of star formation rate
density versus lookback time based on the models and observational data of
\cite{Som}, using a cosmology with $H_0=70$,
$\Omega_M=0.3$, and $\Omega_\lambda=0.7$.  The redshift corresponding
to a particular lookback time is indicated along the top axis of the figure.  
This type of graph is shown at every meeting
pertaining to cosmological star formation, but usually in the
form ``SFR vs. redshift'', which, in my opinion, obscures its utility.  
By plotting SFR vs. linear time, one can integrate by eye
under the curve to obtain the fraction of stars formed, the
fraction of metals produced, or the fraction of gas consumed
since a particular epoch.  For instance, in the left panel,
the hatched region shows that 28\% of the stars and metals in the 
present universe have formed since $z=1$, in the last 8 Gyr.
The hatched region in the right panel shows that 85\% of the stars 
and metals have formed since $z=3$, in the last 10.3 Gyr.
Such estimates serve as starting points to inform our
investigation of the star formation and metal enrichment 
process within {\it individual} galaxies over these epochs. 
While models and computer simulations have provide predictions
for the evolutionary rate of star formation and metal enrichment
in the universe \cite{Pei}\cite{Cen}\cite{Nagamine}, observational 
probes of capable of testing these predictions within
individual galaxies are just now becoming available.   

\begin{figure}
  \includegraphics[height=2.5in,angle=0]{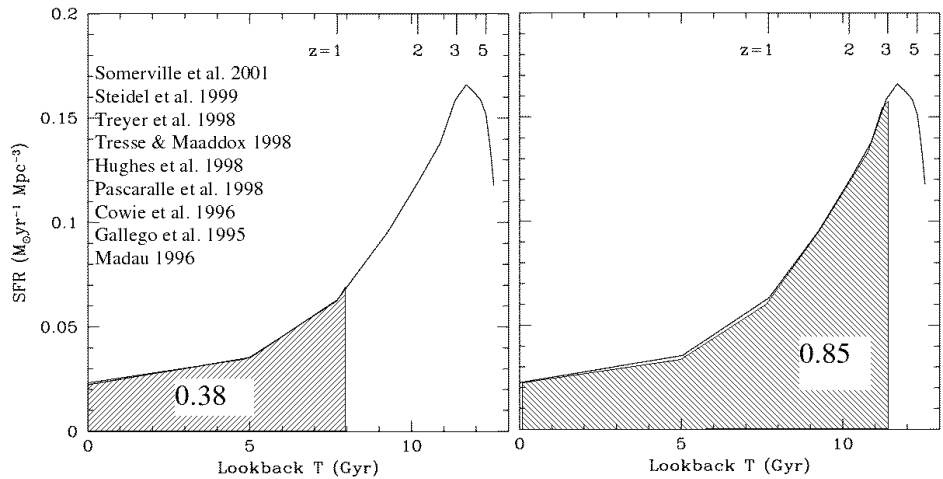}
 \caption{Plot of cosmic star formation rate density versus linear lookback
time. }
\end{figure}

\section{Star Formation Rates and Extinctions}

Measurements of star formation rates in distant galaxies have been
based primarily upon restframe ultraviolet ($L_{1500}$ or $L_{2800}$)
luminosities observed in the optical
\cite{Madau}\cite{Sawicki}\cite{Pascarelle}\cite{Steidel}.  Implied
star formation rates for $z>$1 galaxies range from $\sim10$
M$_\odot~yr^{-1}$ to over 100 M$_\odot~yr^{-1}$ for galaxies which
have luminosities several magnitudes brighter than $L*$.  However,
debate about the amount of extinction internal to these high redshift
galaxies rendered the derived star formation rates uncertain by
factors of several.  More recently, it has become possible to use
restframe optical emission lines observed in near-infrared windows to
make independent estimates which are less affected extinction
\cite{Teplitz}\cite{Pettini98}\cite{KK00}\cite{Pettini01}.  Figure~3
shows one such measurement of a galaxy at $z=2.3$ using the Keck
Telescope+NIRSPEC spectrograph \cite{KK00}.  The H$\alpha$ emission
line is seen in the K-band, the $H\beta$ and
[O~III]$\lambda\lambda$4959,5007 lines are seen in the H-band, and the
[O~II]$\lambda$3727 line is a marginal detection in the J-band.
Strong night sky airglow lines swamp the signal from distant galaxies
and make this type of measurement challenging. Targets must be chosen
to lie at redshifts where the strong diagnostic emission lines fall
between airglow lines, and spectral resolutions exceeding R=2000 are
desirable to separate terrestrial atmospheric features from nebular
lines.
  
\begin{figure}
  \includegraphics[height=5.5in,angle=0]{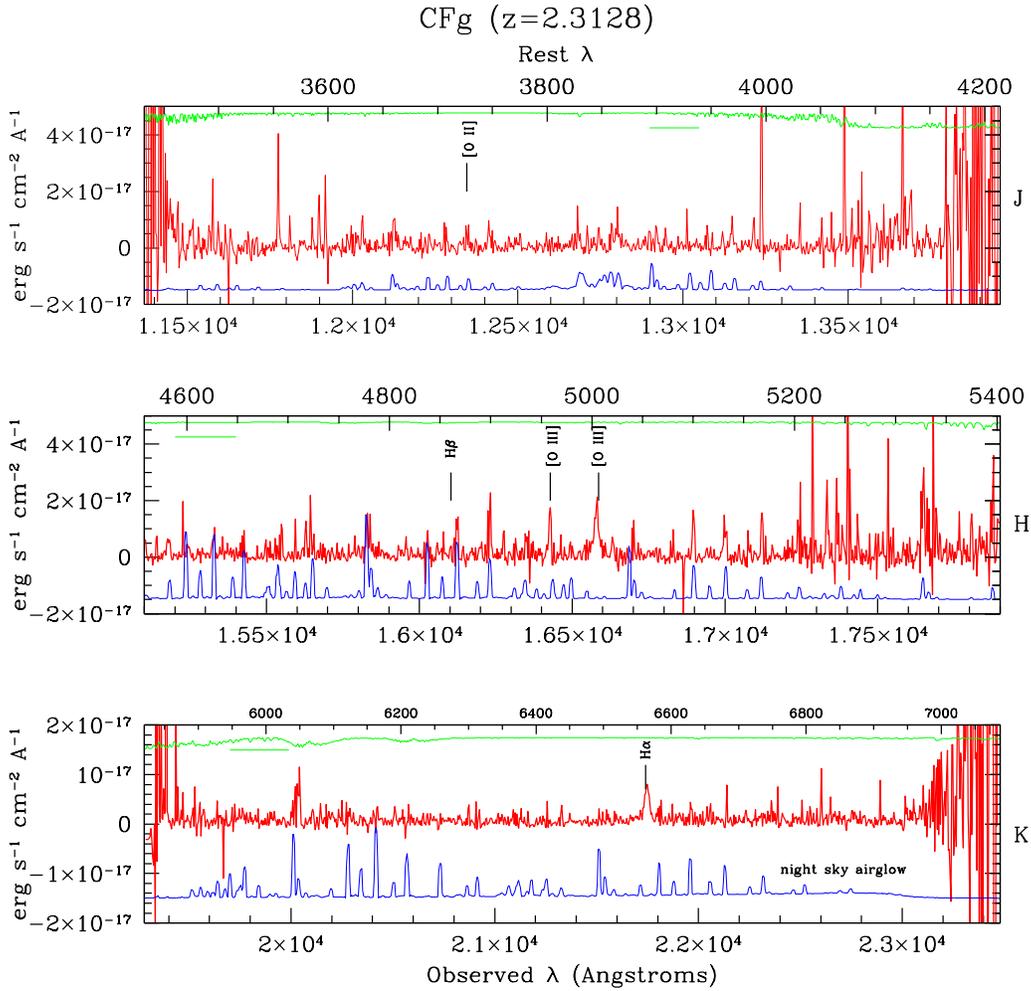}
 \caption{Example of an infrared spectrum showing the optical restframe
emission lines from a $z=2.3$ galaxy \cite{KK00}.   }
\end{figure}

Despite these difficulties, measurements of the H$\alpha$ luminosities
of high redshift galaxies provide an independent estimate of the star
formation rates.  Comparisons show that the SFRs derived from H$\alpha$
luminosities are systematically higher by factors of 2-3 compared to
those inferred from the ultraviolet continuum luminosities at redshifts
from $z=1-2$
\cite{Glazebrook}\cite{Yan}\cite{Hopkins}
out to $z>3$ \cite{Pettini01}\cite{KK00}.  Extinctions at $H\alpha$
are generally $\sim$1 mag versus 2--4 mag at 1500\AA.  The lack of
correlation between the extinction-corrected UV and $H\alpha$-derived
star formation rates suggests that the extinctions are poorly known,
or that the geometry of the dust and gas relative to the stars leads
to different extinctions toward each \cite{Erb1}.  Galaxies with
fainter observed fluxes appear to have ratios $L_{H\alpha}/L_{1500}$
closer to unity, suggesting that extinction may significantly shape
the spectral energy distributions of high redshift galaxies, i.e.,
higher extinctions lead to more extreme $L_{H\alpha}/L_{1500}$ ratios
and lower observed fluxes \cite{Erb1}.  Deep submillimeter continuum
searches reveal a population of extremely red galaxies galaxies (EROs) with
infrared colors $R-K>6$ and large dust content, suggesting that some,
still debated, fraction of
star formation activity is be hidden from view at restframe optical
wavelengths \cite{Lutz}\cite{Ivison}\cite{Frayer}.

\section{ISM Kinematics}

Velocity-revolved optical and near-infrared spectra of stellar and
interstellar lines high redshift objects reveal the dynamical state
of the galaxies and the impact of star formation on their ISM.  $H\alpha$
rotation curves of spiral galaxies out $z=1$ show a well-developed
Tully-Fisher relationship at 8 Gyr lookback times, although the
relation may be offset by 0--2 mag toward higher luminosities
\cite{Vogt}\cite{Rix}\cite{Simard}\cite{Bohm}\cite{Ziegler}, consistent
with a brightening due to increased star formation at earlier epochs.
However, the degree of luminosity evolution is uncertain and
may be strongly color-dependent. 
At redshift beyond $z=2$, spiral galaxies become very rare and yet to
be studied kinematically.  In the few cases where spatially-resolved emission lines
are observed at $z\sim2$, the implied velocity widths and masses are
80-200 \kms\ and 1--6$\times10^{10}$ $M_\odot$.  However, neither
spatially resolved lines nor 1-D H$\alpha$ and [O~III]$\lambda5007$
emission line widths in $z=2-4$ galaxies show a correlation with
galaxy luminosity, suggesting that the observed kinematics reflect bulk motions
of merging sub-clumps or gas in extended halos of galaxies being assembled 
instead of true rotation curves \cite{Erb1}\cite{KK00}.  The overall
linewidths are larger at $z\sim2$ than at $z\sim3$
\cite{Pettini01}\cite{Erb1}\cite{Erb2}, consistent with the
idea of continuing galaxy growth and more ordered dynamics as a galaxy ages.
Given that recognized morphological types are replaced by an
increasing fraction of mergers, irregulars, and compact galaxies
(``proto-galaxies''?) at $z>1$ \cite{Giavalisco}\cite{Driver},
emission line kinematics probably do
not provide a reliable means of estimating their masses \cite{Pettini01}\cite{Erb1}.  
Star formation regions producing the
ionized gas tracers are unlikely to sample the full gravitational
potential of their host systems.  This effect is observed in local
samples of compact and irregular galaxies where the 1-dimensional
$H\alpha$ profiles can be much more narrow than the neutral HI 21-cm
profiles, indicating that the star formation activity covers only a
small portion of the potential well \cite{Pisano}.

\begin{figure}[ht]
  \includegraphics[height=2.9in,angle=0]{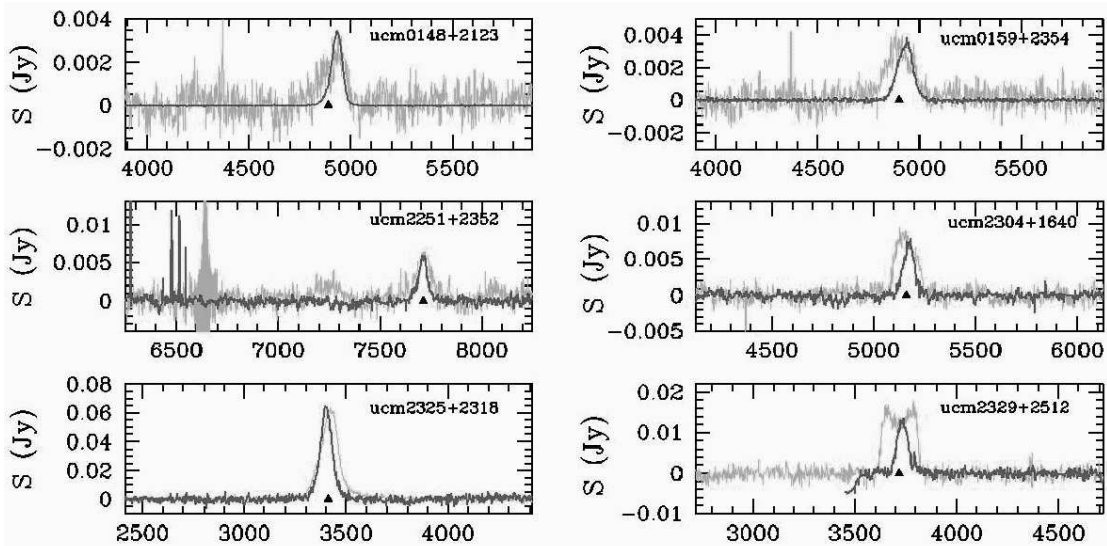}
 \caption{Comparison of ionized gas $H\alpha$ velocity profiles and
neutral hydrogen 21-cm velocity profiles in local irregular and blue
compact galaxies where the star formation traces only a fraction of
the gravitational potential.  H$\alpha$ linewidths provide, therefore, only
a lower limit on the dynamical mass \cite{Pisano}.   }
\end{figure}

Spectra of high-redshift galaxies also reveals systematic differences
between the velocities of the cool or hot ambient ISM, ionized
star-forming ISM, and Ly~$\alpha$ emitting regions, where present.
Species seen in absorption against the starburst continuum are
systematically blue-shifted by 100--500
\kms\ compared to the H$\alpha$ and other nebular tracers of the star
formation regions
\cite{Pettini01}\cite{Erb1}\cite{Erb2}.
This is interpreted as a signature of galactic-scale outflows powered
by stars and supernovae, such as those seen in more local starbursts
(e.g., \cite{Marlowe}\cite{Johnson}\cite{Heckman}\cite{Martin}).  The
redshift of the Ly~$\alpha$ line relative to the starburst region is
understood to be a radiative transfer effect whereby the photons on
the blue wing of the line are preferentially absorbed and scattered by
neutral hydrogen leading to an asymmetric line profile, an effect also
observed in local galaxies \cite{Kunth}.  The impact of dust and
extinction on the inferred internal kinematics of galaxies is also
largely unknown.

\section{Chemical Compositions}

Direct measurements of metallicities\footnote{Here I mean primarily
the {\it nebular oxygen abundance} as tracer of overall gas-phase {\it
metallicity}.} in distant galaxies are now becoming routine, spurred
on by larger telescopes and more capable spectrographs used in deep
galaxy surveys.  Classical nebular diagnostic techniques are applied
to emission lines from star forming galaxies to estimate a
globally averaged gas-phase oxygen abundance using the ratios of
strong H$\beta$, [O~II]$\lambda$3727, and [O~III]$\lambda5007$
emission lines \cite{KKP}.  Figure~4 shows a 3600 s Keck spectrum
and an HST F814W broadband optical image of a starforming galaxy at $z=0.64$
from the DEEP Groth Strip Survey \cite{Ke03}.  This galaxy might be considered
typical of those at similar redshift, having only the strongest emission lines
measurable, and having an irregular appearance, with a possible companion.  

\begin{figure}
  \includegraphics[height=1.8in,angle=0]{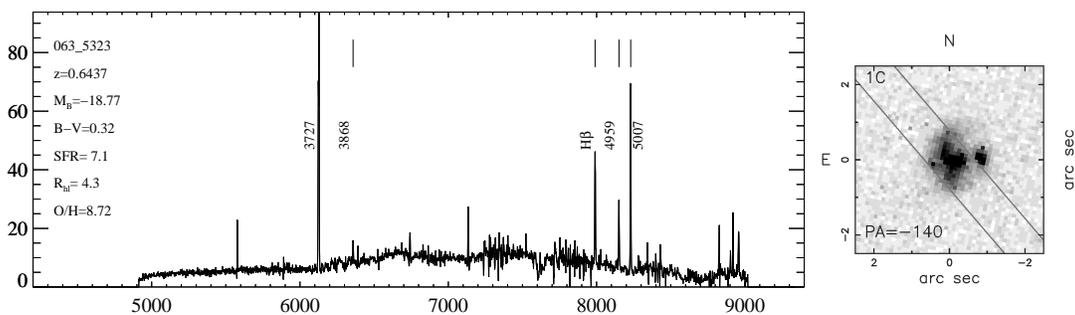}
 \caption{A sample spectrum of a star forming galaxy
at $z=0.64$ showing the prominent nebular emission lines
used for chemical analysis, along with an optical HST-F814W
image  \cite{Ke03}.   }
\end{figure}

Pioneering surveys at $0.1<z<0.98$
\cite{KZ}\cite{CL} initially indicated that intermediate redshift
field and compact starburst galaxies followed the same correlation
between luminosity and metallicity (i.e., the L-Z relation) observed
in local samples (e.g., \cite{Faber}\cite{Lequeux}
\cite{SKH}).  However for luminous $M_B<-21$ objects 
at redshifts $z>2$, near-infrared nebular spectroscopy of the
rest-frame optical emission lines suggested oxygen abundances between
1/3 and 2.0 times solar\footnote{12+log(O/H): The logarithmic
abundance by number of oxygen relative to hydrogen, where the solar
value is 12+log(O/H)= 8.7 \cite{Allende}. The oxygen abundance is
related to, but distinct from $Z$, the mass fraction of metals, often
called the metallicity.  $Z_\odot\simeq0.015$ for the standard solar
abundance distribution.  } ($8.3<12+log(O/H)<9$), placing them 2-4
magnitudes more luminous than $z=0$ galaxies with comparable
metallicities and making them inconsistent with the local L-Z relation
\cite{KK00}\cite{Teplitz}\cite{Pettini01}\cite{Shapley04}. 
Samples studied to date include only the most luminous objects at
these redshifts, and inclusion of fainter objects in future surveys is
needed to tell the whole story.  It is noteworthy that the oxygen
abundances and [N~II]/$H\alpha$ ratios of these $z>2$ objects are
nearly solar at ages $<2$ Gyr.  These initial results demonstrate that
the most luminous (and massive massive?) objects have formed and
retained enough metals during the early stages of galaxy formation to
enrich their mean ISM metallicity to nearly solar values.  Rapid
chemical enrichment is also observed in the central regions of
high-redshift QSOs \cite{Wampler}\cite{Hamann1}\cite{Hamann2}, implying that
only a few generations of massive stars are sufficient to greatly enrich the ISM
provided that the metals are retained within the gravitational potential wells.

\begin{figure}[ht]
  \includegraphics[height=4.0in,angle=0]{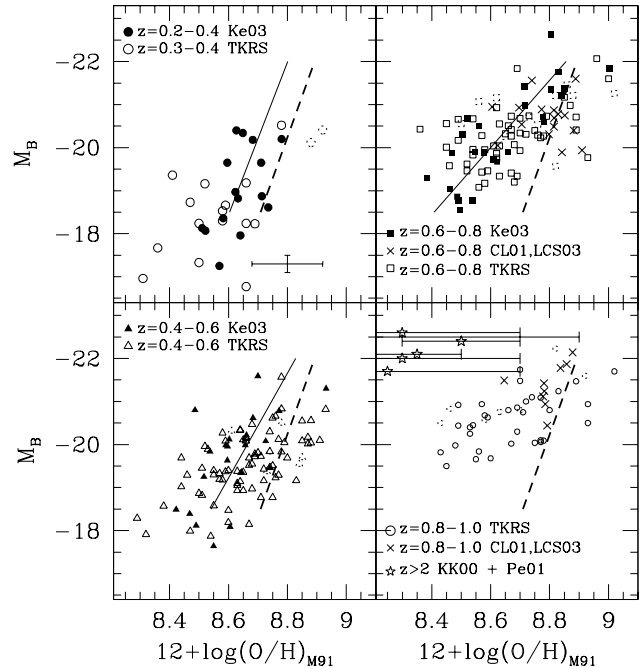}
 \caption{The evolution of the luminosity-metallicity relation
for star forming galaxies as a function of redshift.  
Galaxies at $z>2$ are 2-4 magnitudes more luminous than
local galaxies of comparable metallicity, and deviate greatly
from the local L-Z relation \cite{KK04}.   }
\end{figure}

Evidence for evolution of the L-Z relation with epoch, particularly
among sub L* galaxies with $M_B$ fainter than $-20.5$, grew on the
strength of metallicity measurements in $0.26<z<0.82$ field galaxies
from the Groth Strip Survey (DGSS) \cite{Ke03}, the Canada-France
Redshift Survey (CFRS) fields at $0.47<z<0.92$ \cite{LCS}, the Calar
Alto Deep Imaging Survey (CADIS)
\cite{Maier}, and $0.3<z<1$ in the GOODS-North field \cite{KK04}.  Figure~6 shows the
mean L-Z relations for galaxies in four redshift bins out to $z=1$
using data collected from a variety of surveys.  At the
high-luminosity, metal-rich end of the relation there is much overlap
between distant samples and local galaxies.  At the faint, metal-poor
end of the relation, there appears to be a more dramatic offset from
the local relation. This relatively greater evolution of
low-luminosity galaxies may be evidence for delayed formation epoch
compared to massive galaxies
\cite{Ke03}\cite{Ziegler}\cite{Kodama}.   Also plotted are several
$z>2$ galaxies which depart from the local L-Z relation by several
magnitudes.

\begin{figure}[ht]
  \includegraphics[height=3.6in,angle=0]{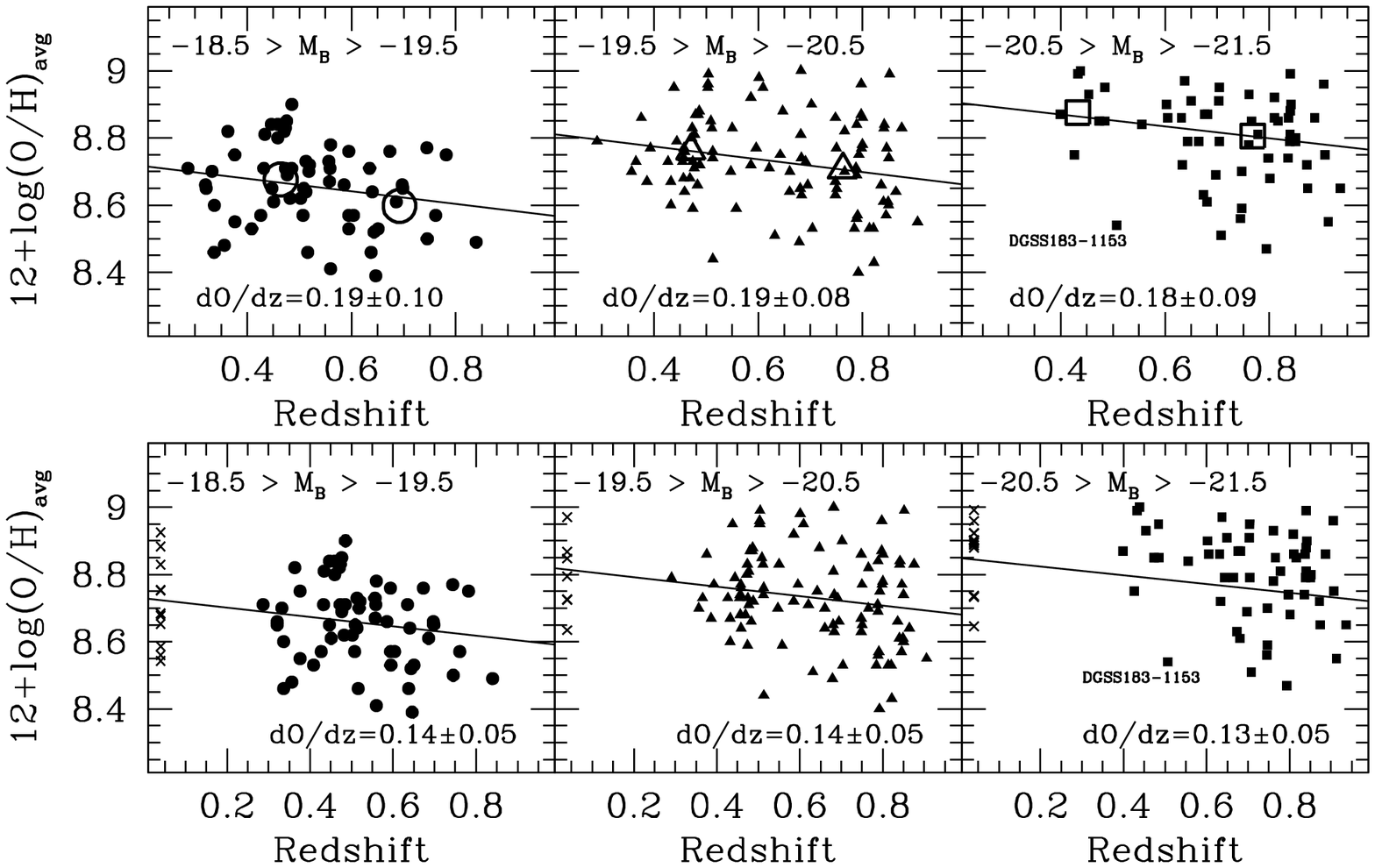}
 \caption{Relation between redshift and gas phase oxygen abundance for 
star forming galaxies over the range $0<z<1$ \cite{KK04}.
   }
\end{figure}

\begin{figure}[ht]
  \includegraphics[height=4.0in,angle=0]{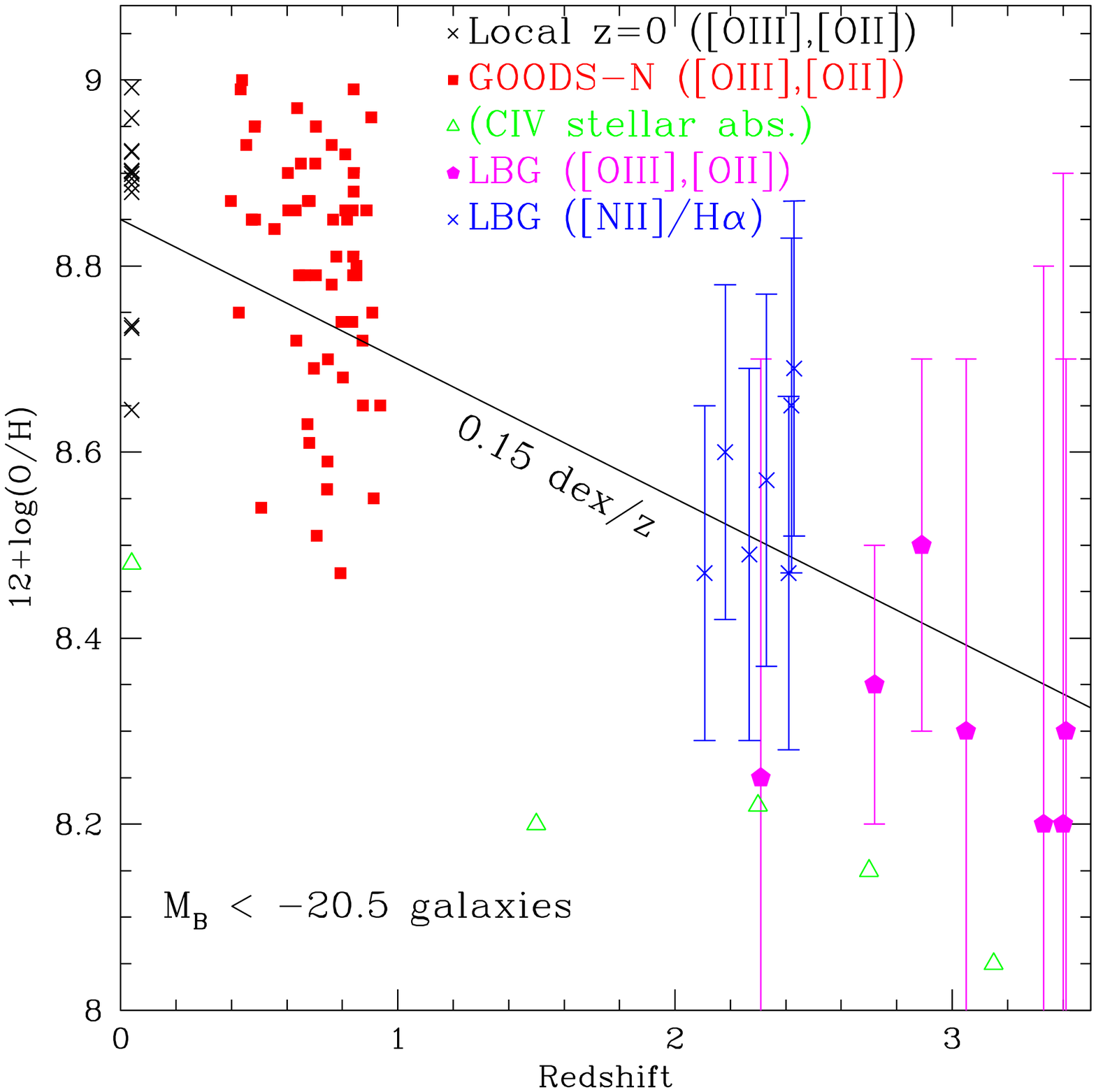}
 \caption{Relation between redshift and gas phase oxygen abundance for galaxies
over the redshift range $0<z<3$ using data from 
\cite{KK04}\cite{Ke03}\cite{Pettini01}\cite{Mehlert}\cite{Shapley04}.  
   }
\end{figure}

Figure~7 provides another look at the chemo-luminous changes in star
forming galaxies by plotting metallicity versus redshift for three
different luminosity bins where there are sufficient numbers of
objects observed over a range of redshifts.  The upper row shows only
objects $0.3<z<1.0$ while the lower row includes local $z=0$ galaxies.
A linear fit to the data over the range $z=0-1$ indicates a metal
enrichment rate of 0.14 dex per unit redshift.  This rate is
remarkably consistent with the rate of oxygen enrichment among
Galactic disk F and G dwarf stars over the same cosmic interval
\cite{Reddy}.  While the star formation and luminosity history of the
Milky Way is not well known, and while the dispersion in oxygen
abundances for Galactic stars is large, the chemical enrichment
process that occurred in the disk of the Milky Way appears to be a
good representation of the chemical enrichment process in the bulk of
the star-forming galaxies over the last 8 Gyr.

Figure~8 demonstrates that the mean rate of metal enrichment in
luminous $M_B<-20.5$ star forming galaxies over the last 11 Gyr
from $z=3$ to $z=0$ is 0.15 dex/z, nearly identical with
rate observed at lower redshift.  
(Note that the slope in this diagram could represented by the unfortunate,
but amusing, nomenclature $\Delta{Z}/\Delta{z}$).  
This enrichment rate is much less dramatic than the model predictions
of 0.3---1 dex/redshift unit \cite{Pei}\cite{Cen}\cite{Nagamine},
but is more nearly consistent with the rates of 0.18---0.26$\pm0.07$
dex/redshift unit seen in damped Ly~$\alpha$ absorption systems 
\cite{Prochaska}\cite{Kulkarni}. \cite{Cen} point out that the
overall level of metal enrichment, and the rate of enrichment with 
time should be functions of local density, with the highest
density objects (i.e., galaxies) reaching metal saturation
very rapidly at earlier times and less dense regions (i.e.,
damped Ly~$\alpha$ and other metal absorption systems) 
enriching more slowly.  Efforts to measure {\it stellar}
metallicities in distant galaxies using photospheric absorption lines
have just begin, but look promising \cite{Mehlert}.  
Figure~8 shows that the
overall level of metal enrichment as determined from
the stellar photospheric C IV absorption lines (triangles; \cite{Mehlert})
is systematically 0.2-0.4 dex lower at all redshifts than the determinations from
nebular lines, suggesting a possible calibration uncertainty
between the methods.  The rate of metal enrichment is similar, however.  

\section{Future Prospects}

The existence of a metallicity-luminosity correlation for high redshift
galaxies offers the prospect of reliable metallicity estimates for use
in modeling spectral energy distributions
and dust contributions in the distant universe
based on extinction laws and dust properties in nearby galaxies.
High redshift galaxies may not be all that different from
local starbursts in terms of their
metal and dust content, even if the star formation rates
are much larger than all but the most impressive
Ultra Luminous InfraRed Galaxies (ULIRGS) locally.
If mergers are responsible for the dominance of irregular morphologies
as galaxies assemble at high redshift, it will be difficult to
use observed linewidths or what appear to be rotation curves to measure masses.
With adaptive optics and sensitive integral field spectrographs
on the largest current telescopes, it will be possible to map velocity
fields in galaxies out to $z\sim3$ to obtain a better 
dynamical picture.  A true understanding of the fundamental galaxy
correlations (i.e., Tully-Fisher, L-Z) and how they evolve at high redshift
ultimately requires sampling the intrinsically faint sub L* galaxy population 
at $z>2$, a feat that will not be easily accomplished with the current 
generation of telescopes.   



\bibliographystyle{aipprocl} 

{}

\end{document}